\documentstyle[cite]{article}

\def\fun#1#2{\lower3.6pt\vbox{\baselineskip0pt\lineskip.9pt
\ialign{$\mathsurround=0pt#1\hfil##\hfil$\crcr#2\crcr\sim\crcr}}}

\begin{document}

\title{THE CONCEPT OF MASS IN THE EINSTEIN YEAR
\footnote{Presented at the 12th Lomonosov conference
on Elementary Particle Physics, Moscow State University, August 25-31.}}
\author{L.B. Okun
 \\ \small State Research Center, Institute for Theoretical and
Experimental Physics, \\ \small 117218, Moscow, Russia} 
\maketitle
\abstract{ Various facets of the concept of mass are discussed.
The masses of elementary particles and the search for higgs. The
masses of hadrons. The pedagogical virus of relativistic mass.}

\section{From ``Principia'' to Large Hadron Collider (LHC)}

The term ``mass' was introduced into mechanics by Newton in 1687
in his ``{\it Principia}'' \cite{1}. He defined it as the amount
of matter. The generally accepted definition of matter does not
exist even today. Some authors of physics text-books do not
consider photons -- particles of light -- as particles of matter,
because they are massless. For the same reason they do not
consider as matter the electromagnetic field. It is not quite
clear whether they consider  as matter almost massless neutrinos,
which usually move with velocity close to that of light. Of course
it is impossible to collect a handful of neutrinos similarly to a
handful of coins. But in many other respects both photons and
neutrinos behave like classical particles, while the
electromagnetic field is the basis of our understanding of the
structure of atoms. On the other hand, the so-called weak bosons
$W^+$, $W^-$, $Z^0$ are often not considered as particles of
matter because they are too heavy and too short-lived.

Even more unusual are such particles as gluons and quarks. Unlike
atoms, nucleons, and leptons, they do not exist in a free state:
they are permanently confined inside nucleons and other hadrons.

There is no doubt that the problem of mass is one of the key
problems of modern physics. Though there is no common opinion even
among the experts what is the essence of this problem. For most of
particle theorists, as well as members of LHC community, the
solution of the problem is connected with the quest and discovery
of the higgs -- scalar boson which in the Standard Model is
responsible for the masses of leptons and quarks and their
electroweak messengers: $W$ and $Z$. The discovery of higgs and
the study of higgs sector might elucidate the problem of the
pattern of hierarchy of masses of leptons and quarks: from milli
electron Volts for neutrinos to about 180 GeV for $t$-quark. For
many physicists it is a QCD problem: how light quarks and massless
gluons form massive nucleons and atomic nuclei. Still for majority
of confused students and science journalists there is no
difference between mass of a body $m$ and its energy $E$ divided
by $c^2$: they believe in the ``most famous formula $E = mc^2$''.

If higgs exists, its discovery will depend on the funding of the
particle physics. In 1993 the termination of the SSC project sent
the quest for the higgs into a painful knockdown.  The decision
not to order in 1995 a few dozen of extra superconducting cavities
prevented, a few years later, LEP II from crossing the 115 GeV
threshold for the mass of the higgs.

If we are lucky and higgs is discovered around year 2010 at LHC,
then the next instrument needed to understand what keeps the
masses of the higgs below 1 TeV scale, is ILC (International
linear collider). This machine would provide a clean environment
for the study of higgs production and decays. It could also be
used for discovery and study of light  supersymmetric particles
(SUSY). A prototype of ILC was suggested a few years ago by DESY
as the project TESLA. There was no doubt that if funded, TESLA
would work, but the funding was not provided by the German
government. The new variant of ILC envisions increasing the
maximal center of mass energy of colliding electron and positron
from ) 0.5 TeV to 1 TeV. If everything goes well, ILC can start
before 2020.

Further increase of energy, to say, 5 TeV, would call for a
machine of the type of CLIC (Compact linear collider) the project
of which is under discussion at CERN for more than a decade. In
this machine the role of clystrons is supposed to play a low
energy but very high current ``decelerator'' the energy of which
would be pumped into the high energy accelerator part of CLIC.
Unlike situation with ILC, even the mere feasibility of CLIC is
not clear now. Special experimental research to ascertain the
feasibility is going on at CERN.

The discussion of higgses, neutrinos and QCD in connection with
the fundamental problems of mass is often accompanied and even
overshadowed by a ``pseudoproblem'' of the so-called
``relativistic mass'' (see section 5).

\section{Mass in Newtonian Mechanics}

The more basic is a physical notion, the more difficult to define
it in words. A good example give the 1960s editions of
``Encyclopedia Britannica'' where energy is defined in terms of
work, while the entry ``work'' refers to labour and professional
unions. Most people have intuitive notions of space and time.
Every physicist has intuitive notions of energy, mass, and
momentum. But practically everybody has difficulties in casting
these notions into words without using mathematics.

Though the definition of mass (``Definition I: The  quantity of
matter is the measure of the same, arising from its density and
bulk conjointly'') given by Newton in his ``{\it Principia}''
\cite{1} was so unclear that scholars are discussing its logical
consistency even today, the equations of Newtonian mechanics are
absolutely self-consistent. Mass $m$ enters in the relations of
velocity ${\rm \bf v}= d{\rm \bf r}/dt$ and momentum $\rm\bf p$:
\begin{equation}
{\rm \bf p} = m {\rm \bf v} \;\; , \label{1}
\end{equation}
as well as acceleration ${\rm \bf a} = d{\rm \bf v}/dt$ and force
${\rm\bf F}$:
\begin{equation}
{\rm\bf F} = d{\rm\bf p}/dt =m{\rm\bf a} \;\; . \label{2}
\end{equation}

It also enters in the equation defining the force of gravity with
which a body with mass $m_1$ at point ${\rm\bf r}_1$ attracts
another body with mass $m_2$ at point ${\rm\bf r}_2$:
\begin{equation}
{\rm\bf F}_g = -G m_1 m_2 {\rm\bf r}/r^3 \;\; . \label{3}
\end{equation}
Here ${\rm\bf r} = {\rm\bf r}_2 -{\rm\bf r}_1$, $r =|{\rm\bf r}|$,
while $G$ is the famous Newton constant:
\begin{equation}
G = 6.67 \cdot 10^{-11} {\rm m}^3 {\rm kg}^{-1} {\rm s}^{-2} \;\;
. \label{4}
\end{equation}

The kinetic energy of a body is defined as
\begin{equation}
E_k = {\rm\bf p}^2/2m = m{\rm\bf v}^2/2 \;\; . \label{5}
\end{equation}
The potential gravitational energy:
\begin{equation}
U_g = -Gm_1 m_2 /r \;\; , \label{6}
\end{equation}
while the total energy in this case is
\begin{equation}
E=E_k +U_g \;\; . \label{7}
\end{equation}

The total energy is conserved. When a stone falls on the earth,
its potential energy decreases (becomes more negative), kinetic
energy increases: so that the total energy does not change. When
the stone hits the ground, its kinetic energy is shared by the
ambient molecules raising the local temperature.

One of the greatest achievements  of the XIX century was the
formulation of the laws of conservation of energy and momentum in
all kinds of processes.

At the beginning of the XX century it was realized that
conservation of energy is predetermined by uniformity of time,
while conservation of momentum -- by uniformity of space.

But let us return to the notion of force. People strongly felt the
force of gravity throughout the history of mankind, but only in
XVII century the equations (\ref{3}) and (\ref{6}) were
formulated.

An important notion in this formulation is the notion of
gravitational potential $\varphi_g$. The gravitational potential
of a body with mass $m_1$ is
\begin{equation}
\varphi_g = -\frac{Gm_1}{r} \label{8'}
\end{equation}

Thus the potential energy of a body with mass $m_2$ in a potential
(\ref{8'}) is
\begin{equation}
U_g = m_2 \varphi_g \;\; , \label{9'}
\end{equation}
which coincides with eq. (\ref{6}).

A century later similar equations were formulated for another
long-range interaction, the electrical one:
\begin{equation}
{\rm\bf F}_e = e_1 e_2 {\rm\bf r}/r^3 \;\; , \label{8}
\end{equation}
\begin{equation}
U_e = e_1 e_2/r  \;\; , \label{9}
\end{equation}
\begin{equation}
\varphi_e = e_1/r \;\; , \label{12'}
\end{equation}
\begin{equation}
U_e = e_2 \varphi_e \;\; . \label{13'}
\end{equation}

In these equations, which define the Coulomb force, Coulomb
potential energy, and Coulomb potential respectively, $e_1$ and
$e_2$ are electrical charges of two bodies.

An important role in the theory of electricity is played by the
strength of electric field ${\rm\bf E}$. Charge $e_1$ creates
field with strength
\begin{equation}
{\rm\bf E} = e_1 {\rm\bf r}/r^3 \;\; . \label{14'}
\end{equation}
Thus
\begin{equation}
{\rm\bf F}_e = e_2 {\rm\bf E} \;\; \label{15'}
\end{equation}

As most of matter around us is electrically neutral, the
electrical interaction was known for centuries only as a kind of
trifle. Unlike mass $m$, which is always positive, the charge $e$
has two varieties: positive and negative. Two charges of opposite
sign attract each other, while those of the same sign are
repelling. Protons residing in the nucleus of an atom have
positive charge, the charge of electrons, which form atomic
shells, is negative. As a result the atom is electrically neutral.

Electrical interaction and its ramifications determine the main
features of atoms, molecules, condensed matter, and biological
cells. Gravitational interaction is too feeble to play any role at
that level. To see this consider an electron and a proton. Their
masses are
\begin{equation}
m_e = 9.1 \cdot 10^{-31} \; {\rm kg} \;\; , \label{10}
\end{equation}
\begin{equation}
m_p = 1.7 \cdot 10^{-27} \; {\rm kg} \;\; . \label{11}
\end{equation}
Their electric charges:
\begin{equation}
e_p = -e_e = e = 4.8 \cdot 10^{-10} \; {\rm esu} \;\; , \label{12}
\end{equation}
where esu denotes electrostatic unit:
\begin{equation}
(1 \; {\rm esu})^2 = 10^{-9} \; {\rm kg \; m}^3 {\rm s}^{-2} \;\;
. \label{13}
\end{equation}
Hence
\begin{equation}
-e_e e_p = 2.3 \cdot 0^{-28} \; {\rm kg \; m}^3 {\rm s}^{-2} \;\;
. \label{15}
\end{equation}
On the other hand
\begin{equation}
G m_e m_p = 1.03 \cdot 10^{-67} \; {\rm kg \; m}^3  {\rm s}^{-2}
\;\; , \label{14}
\end{equation}

Thus in an atom the gravitational force is $\sim 10^{-40}$ of
electric one.

The importance of gravity for our every day life is caused by the
huge number of atoms in the  earth, and hence by its very large
mass:
\begin{equation}
M = 5.98 \cdot 10^{24} \; {\rm kg} \;\; . \label{16}
\end{equation}
Taking into account the value of the radius of the earth
\begin{equation}
R= 6.38 \cdot 10^6 \; {\rm m} \;\; , \label{17}
\end{equation}
we find the gravitational force of the earth acting on a body with
mass $m$ close to its surface:
\begin{equation}
{\rm\bf F}_g = m \;{\rm\bf g} \;\; , \label{18}
\end{equation}
where ${\rm\bf g}$ is acceleration directed towards the center of
the earth:
\begin{equation}
g = |{\rm\bf g}| =\frac{6.67 \cdot 10^{-11} \cdot 5.98 \cdot
10^{24}}{(6.38 \cdot 10^6)^2} {\rm m \; s}^{-2} = 9.8 \; {\rm m \;
s}^{-2} \;\; . \label{19}
\end{equation}

Let us note that for gravity ${\rm \bf g}$ plays the role of the
strength of gravitational field, which is analogous to the role of
${\rm\bf E}$ for electricity. The acceleration ${\rm \bf g}$ does
not depend on the mass or any other properties of the attracted
body. In that sense it is universal. This universality was
established by Galileo early in the 17th century in his
experiments with balls rolling down an inclined plane. One can see
this plane, with little bells ringing when a ball passes them, in
Florence. (Apocryphal history tells that Galileo had discovered
universality of $g$ by dropping balls from the Tower of Pisa.)

Gravitational and electric interactions are the only long-range
interactions the existence of which has been established. Two
other fundamental interactions, referred to as strong and weak,
have very short ranges: $10^{-15}$ m and $10^{-18}$ m
respectively. Their first manifestations were discovered about a
century ago in the form of radioactivity. Their further study has
lead to new disciplines: nuclear physics and the physics of
elementary particles.

Quite often you might find in the literature, in the discussion of
Newtonian mechanics, the terms ``inertial mass'' $m_i$ and
``gravitational mass'' $m_g$. The former is used in equations
(\ref{1}), (\ref{2}), (\ref{5}), defining the inertial properties
of bodies. The latter is used in equations (\ref{3}), (\ref{6}),
(\ref{18}), describing the gravitational interaction. After
introducing these terms a special law of nature is formulated:
\begin{equation}
m_i = m_g \;\; , \label{20}
\end{equation}
which is called upon to explain the universality of $g$.

However Galileo had discovered this universality before the notion
of mass was introduced by Newton, while from Newton equations
(\ref{1}), (\ref{2}), (\ref{3}) the universality of $g$ follows
without additional assumptions. Thus the notions and notations
$m_i$ and $m_g$ are simply redundant. As we will see later, their
introduction is not only redundant, but contradicts the General
Theory of Relativity, which explains why the same mass $m$ enters
equations (\ref{1}) - (\ref{3}).

The advocates of $m_i$ and $m_g$ argue by considering the
possibility that in the future the more precise experimental tests
might discover a small violation of Galilean universality. But
that would mean that a new feeble long-range force exists in
nature. In literature this hypothetical force is often referred to
as a ``fifth force'' (in addition to the four established ones).
When and if the ``fifth force'' is discovered it should be
carefully studied. But at present it should not confuse the
exposition of well established physical laws. Especially confusing
and harmful are $m_i$ and $m_g$ in the text-books for students.

At that point it is appropriate to summarize the properties of
mass in Newtonian mechanics:

\begin{enumerate}
\item Mass is a measure of the amount of matter.
\item Mass of a body is a measure of its inertia.
\item Masses of bodies are sources of their gravitational
attraction to each other.
\item Mass of a composite body is equal to the sum of masses of
the bodies that constitute it; mathematically that means that mass
is additive.
\item Mass of an isolated body or isolated system of bodies is
conserved: it does not change with time.
\item Mass of a body does not change in the transition from one
reference frame to another.
\end{enumerate}

\section{Mass in Special Relativity}

Of great conceptual importance in modern science is the principle
of relativity first stated by Galileo: A rectilinear  motion of a
physical system with constant velocity relative to any external
object is unobservable within the system itself. The essence of
this principle was beautifully exposed in the famous book
``Dialogue Concerning the Chief World Systems -- Ptolemaic and
Copernican'', published in 1632 \cite{2}:

``Shut yourself up with some friend in the main cabin below decks
on some large ship, and have with you there some flies,
butterflies, and other small flying animals. Have a large bowl of
water with some fish in it; hang up a bottle that empties drop by
drop into a wide vessel beneath it. With the ship standing still,
observe carefully how the little animals fly with equal speed to
all sides of the cabin. The fish swim indifferently in all
directions; the drops fall into the vessel beneath; and, in
throwing something to your friend, you need throw it no more
strongly in one direction than another, the distances being equal:
jumping with your feet together, you pass equal spaces in every
direction. When you have observed all these things carefully
(though there is no doubt that when the ship is standing still
everything must happen in this way). Have the ship proceed with
any speed you like, so long as the motion is uniform and not
fluctuating this way and that, you will discover not the least
change in all the effects named, nor could you tell from any of
them whether the ship was moving or standing still. In jumping,
you will pass on the floor the same spaces as before, nor will you
make larger jumps toward the stern than toward the prow even
though the ship is moving quite rapidly, despite the fact that
during the time you are in the air the floor under you will be
going in a direction opposite to your jump. In throwing something
to your companion, you will need no more force to get it to him
whether he is in the direction of the bow or the stern, with
yourself situated opposite. The droplets will fall as before into
the vessel beneath without dropping toward the stern, although
while the drops are in the air the ship runs many spans. The fish
in their water will swim toward the front of their bowl with no
more effort than toward the back, and will go with equal ease to
bait placed anywhere around the edges of the bowl. Finally the
butterflies and flies will continue their flights indifferently
toward every side, nor will it ever happen that they are
concentrated toward the stern, as if tired out from keeping up
with the course of the ship, from which they will have been
separated during long intervals by keeping themselves in the air.
And if smoke is made by burning some incense, it will be seen
going up in the form of a little cloud, remaining still and moving
no more toward one side than the other. The cause of all these
correspondences of effects is the fact that the ship's motion is
common to all the things contained in it, and to the air also.
That is why I said you should be below decks; for if this took
place above in the open air, which would not follow the course of
the ship, more or less noticeable differences would be seen in
some of the effects noted.''

Sometimes one can hear that the ship of Galileo was discussed two
centuries earlier by cardinal Nicolaus Cusanus (1401 -- 1464) in
his book ``De docta ignorata'' (``On the scientific ignorance'')
published in 1440. Indeed, one can read in volume II, at the
beginning of chapter XII ``The properties of the earth'':

``It is clear to us that the earth is actually moving, though we
do not see this, as we feel the movement only through comparison
with a point at rest. Somebody on a ship in the middle of waters,
without knowing that water is flowing and without seeing the
shores, how could he ascertain that  the ship is moving?''
\cite{3} \footnote{I am grateful to Peter Zerwas for arousing my
interest to Nicolaus Cusanus.}. The relative character of motion
is expressed in these lines quite clearly. But the cabin of
Galileo's ship is full of various phenomena and experiments,
proving that observable effects look the same in any inertial
reference frame. At this point we define an inertial reference
frame, as that which moves rectilinearly with constant velocity
with respect to the stars. We shall give a more accurate
definition when considering General Theory of Relativity.

If the velocity of the ship is ${\rm \bf u}$ and it moves along
the axis $x$, the coordinates of two inertial frames are connected
by equations:
\begin{equation}
t^{\prime} = t \;\; ,  \label{21}
\end{equation}
\begin{equation}
x^{\prime} = x+ ut \;\; ,  \label{22}
\end{equation}
\begin{equation}
y^{\prime} = y \;\; , \;\;  z^{\prime} = z \;\; , \label{23}
\end{equation}
where $u = |{\rm \bf u}|$, the primed coordinates refer to the
shore, while unprimed to the ship. From the definition of velocity
${\rm \bf v} = d{\rm\bf r}/dt$ one easily sees that
\begin{equation}
{\rm \bf v}^{\prime} = {\rm \bf v} + {\rm \bf u} \label{24}
\end{equation}
and that  ${\rm \bf v}^{\prime}$, ${\rm \bf p}^{\prime}$, ${\rm
\bf a}^{\prime}$, ${\rm \bf F}^{\prime}$, ${\rm \bf F}_g^\prime$,
$E_k^\prime$, $U_g^{\prime}$, ${\rm \bf F}_e^\prime$, $U_e^\prime$
satisfy the same equations (\ref{1}) - (\ref{9}) as their unprimed
analogues.

Galilean principle of relativity is the quintessence of Newtonian
mechanics. Nevertheless the latter is called non-relativistic
mechanics, as opposed to Einsteinian mechanics which is called
relativistic. This is one of many examples of lack of complete
consistency in the language of physics which is a natural product
of its evolution. The point is that Newtonian mechanics satisfies
the Galilean principle of relativity only partially. The cabin of
the original Galilean ship did not contain apparatuses that were
able to measure the velocity of light. This velocity was first
established in 1676 by Danish astronomer O. Roemer (1644 - 1710),
who deduced from the observations of the moons of Jupiter,
performed by J. Cassini, that it is $2.4 \cdot 10^5$ km s$^{-1}$.
Further measurements during three centuries established its
present value: $c =3 \cdot 10^5$ km s$^{-1}$.

Of greatest importance was the discovery made two centuries later
by American physicists A. Michelson and E. Morley. By using a
special two arm rotating interferometer they established that the
velocity of light did not depend on the angle between the light
ray and the vector of velocity of the earth on its journey around
the sun. In this experiment the earth itself played the role of
Galilean ship. That result signalled that the simple law
(\ref{24}) of addition of velocities is not valid for light.

This, in its turn, meant that the coordinate transformations
(\ref{21}) - (\ref{24}) should be changed when $v$ and/or $u$ (due
to relative character of velocity) are of the order of $c$.

This change had been performed by H. Lorentz (1904) \cite{4}, H.
Poincare (1905) \cite{5,6} and A. Einstein (1905) \cite{7,8}.
Lorentz considered deformation of electron moving through the so
called ether, filling all the universe, and introduced  primed
spatial and time coordinates, as purely auxiliary quantities.
Poincare and Einstein wrote transformations between primed and
unprimed coordinates:
\begin{equation}
t^\prime = (t+ux/c^2)\gamma \;\; , \label{25}
\end{equation}
\begin{equation}
x^\prime = (x+ut)\gamma \;\; , \label{26}
\end{equation}
\begin{equation}
y^{\prime} = y \;\; , \;\;  z^{\prime} = z \;\; , \label{27}
\end{equation}
where
\begin{equation}
\gamma = 1/\sqrt{1-u^2/c^2} \;\; . \label{28}
\end{equation}
They were called Lorentz transformations by Poincare and later by
Einstein.

Poincare believed in ether and considered that the remaining
problem is to understand it. Einstein simply dispensed with ether,
he considered transformations (\ref{25}) - (\ref{28}) as a direct
expressions of properties of space and time. Galilean relativity
of inertial motion resulted in relativity of simultaneity, of
time, and of length.

Proceeding from his article \cite{7} Einstein came \cite{8} to a
fundamental conclusion that a body at rest has rest-energy $E_0$:
\begin{equation}
E_0 = mc^2 \;\; . \label{29}
\end{equation}
Here $m$ is the mass of the body, while index $0$ in $E_0$
indicates that this is the energy in the body's rest frame.

In 1906 M. Planck explicitly wrote the expressions of total energy
$E$ and momentum ${\rm \bf p}$ of a body with arbitrary value of
its velocity ${\rm \bf v}$:
\begin{equation}
E = mc^2 \gamma \;\; , \label{30}
\end{equation}
\begin{equation}
{\rm \bf p} = m {\rm \bf v} \gamma \;\; , \label{31}
\end{equation}
where
\begin{equation}
\gamma = (1-v^2/c^2)^{-1/2} \;\;. \label{32}
\end{equation}

These expressions can be easily derived by assuming that $E$ and
${\rm \bf p}$ transform in the same way as $t$ and ${\rm \bf r}$,
each pair $(E, {\rm \bf p}c)$ and $(t, {\rm \bf r}/c)$ forming a
four-dimensional vector. Indeed, by applying Lorentz
transformations to a body at rest, taking into account relation
(\ref{29}) and writing $v$ instead of $u$, we come to (\ref{30}) -
(\ref{32}). Of course, the isotropy of space should be also
accounted for.

The notion of four-dimensional space-time was introduced in 1908
by G. Minkowski \cite{10}. While four-vectors transform under
Lorentz transformations (rotations in Minkowskian pseudo-Euclidian
space), their squares are Lorentz-invariant:
\begin{equation}
\tau^2 = t^2 - ({\rm \bf r}/c)^2 \;\; , \label{33}
\end{equation}
\begin{equation}
m^2 c^4 = E^2 -({\rm \bf p} c)^2 \;\; . \label{34}
\end{equation}
Here $\tau$ is the so-called proper time, while $m$, as before, is
the mass of a body. But now it acquires a new meaning, which was
absent in Newtonian mechanics. (Note that for a body at rest
(${\rm \bf p} =0$) one recovers from eq. (\ref{34}) the relation
(\ref{29}) between mass and rest-energy.)

It is impossible to discuss the concept of mass without explicitly
basing the discussion on the achievements of XX century physics
and especially on the notion of elementary particles such as
electrons, photons and neutrinos, less elementary, such as protons
and neutrons (in which quarks and gluons are confined), or
composite, such as atoms and atomic nuclei. It is firmly
established that all particles of a given kind (for instance all
electrons) are identical and hence have exactly the same value of
mass. The same refers to protons and neutrons.

Atoms and atomic nuclei ask for further stipulations because each
of these composite systems exists not only in its ground state,
but can be brought to one of its numerous excited states (energy
levels). For instance, a hydrogen atom is a bound system of a
proton and electron, attracted to each other by Coulomb force
(\ref{8}). As proton is approximately two thousand times heavier
than electron, one usually speaks about electron moving in the
Coulomb potential (\ref{9}) of proton. According to the laws of
quantum mechanics this movement is quantized, forming a system of
levels. The lowest level is stable, the excited ones are unstable.
Electron jumps from a higher level to a lower one by emitting a
quantum of light - photon. Finally it reaches the ground level.

The energy in atomic, nuclear and particle physics is measured in
electron Volts: 1 eV is the energy which electron gains by
traversing a potential of 1 Volt; $1 \; {\rm keV} = 10^3$ eV, $1
\; {\rm MeV} = 10^6$ eV, $1 \; {\rm GeV} = 10^9$ eV, $1 \; {\rm
TeV} = 10^{12}$ eV, $1 \; {\rm PeV} = 10^{15}$ eV, $1 \; {\rm EeV}
= 10^{18}$ eV.

The binding energy of electron at the ground level of hydrogen
atom is 13.6 eV. Due to relation between rest-energy and mass it
is convenient to use as a unit of mass 1 eV/c$^2$. The mass of
electron $m_e = 0.511$ MeV/c$^2$, the mass of proton $m_p = 0.938$
GeV/c$^2$. The mass of a hydrogen atom in its ground state is by
13.6 eV/c$^2$ smaller than the sum $m_e + m_p$. This mass
difference is often referred to as defect of mass.

As $c$ is a universal constant, it is appropriate to use it in
relativistic physics as a unit of velocity and hence to put $c=1$
in all above values of masses and defects  of mass. In what
follows we will use as units of mass eV and its derivatives: keV,
MeV, GeV, etc.

One eV is a tiny unit when compared with Joule (J) or kilogram:
\begin{equation}
1 \; {\rm J} = 6.24 \cdot 10^{18} \; {\rm eV} \;\; , \;\; 1 \;
{\rm eV} = 1.6 \cdot 10^{-19} \; {\rm J} \;\; , \label{35}
\end{equation}
\begin{equation}
1 \; {\rm kg} = 5.61 \cdot 10^{35} \; {\rm eV} \;\; , \;\; 1 \;
{\rm eV} = 1.78 \cdot 10^{-36} \; {\rm kg} \;\; .\label{36}
\end{equation}
However it is four orders of magnitude larger than one degree of
Kelvin (K).
\begin{equation}
1 \; {\rm K} = 0.86 \cdot 10^{-4} \; {\rm eV} \;\; , \;\; 1 \;
{\rm eV} = 1.16 \cdot 10^{4} \; {\rm K} \;\; \label{37}
\end{equation}
(In eq. (\ref{37}) we put dimensional Boltzmann factor $k$ equal
to unity, taking into account that $kT$, where $T$ is temperature,
characterizes the mean energy of an ensemble of particles.) Let us
estimate the relative change of mass in a few everyday processes.

The light from the sun is absorbed by vegetation on the earth to
produce carbohydrates via reaction of photosynthesis: $$ {\rm
light} + 6{\rm CO}_2 + 6{\rm H}_2{\rm O} = 6{\rm O}_2 + {\rm C}_6
{\rm H}_{12}{\rm O}_6 \;\; .$$ The total energy of light required
to produce one molecule of ${\rm C}_6{\rm H}_{12}{\rm O}_6$ is
about 4.9 eV. This does not mean that the photons are massive.
They are massless, but the kinetic energy of photons is
transformed into the rest energy of carbohydrates.

A combustion of methane in the gas burner of a kitchen stove:
\begin{equation}
{\rm CH}_4 + 2{\rm O}_2 \to {\rm CO}_2 + 2{\rm H}_2{\rm O} \;\; .
\label{38}
\end{equation}
In this reaction 35.6 MJ of heat is released per cubic meter of
methane. Since the density of methane is 0.72 kg/m$^3$ and density
of oxygen is 1.43 kg/m$^3$ $$ \frac{\Delta m}{m} = \frac{35.6
\cdot 6.24 \cdot 10^6 \cdot 10^{18}}{(0.72 + 2 \cdot 1.43) \cdot
5.61 \cdot 10^{35}} = 1.1 \cdot 10^{-10} \;\; , $$ where in the
nominator  eq.(\ref{35}) and in denominator eq.(\ref{36}) are
used.

We can look at this  result differently by starting from Avogadro
number:
\begin{equation}
N_A = 6.022 \cdot 10^{23} \; {\rm mol}^{-1} \label{39}
\end{equation}
and molar volume (for ideal gas) $$ 22.4 \cdot 10^{-3} \; {\rm
m}^3 \cdot {\rm mol}^{-1} \;\; . $$ This means that a cubic meter
of methane contains $2.69 \cdot 10^{25}$ molecules. Thus, burning
of one molecule of methane releases $$ \frac{35.6 \cdot 6.24 \cdot
10^{24}}{2.69 \cdot 10^{25}} = 8.3 \; {\rm eV} $$ Now we estimate
the mass of one molecule of methane plus 2 molecules of oxygen:
$16 \times 5 \cdot 0.94 \; {\rm GeV} = 75$ GeV, and calculate
$\Delta m/m$ at molecular level (we use 0.94 GeV as the mass of a
nucleon): $$8.3 \; {\rm eV}/75 \; {\rm GeV} = 1.1 \cdot 10^{-10}
\;\; . $$ Thus, we see that the sum of masses of molecules on the
right hand side of eq. (\ref{38}) is by 8.3 eV smaller than that
on the left-hand side. This mass difference is exploited in
cooking.

Another example is the melting of ice. It takes $0.334 \cdot 10^6$
J to melt a kilogram of ice. That means that in this case the
relative increase of mass $\Delta m /m$ is (see eqs. (\ref{35})
and (\ref{36})): $$ \Delta m/m = 0.334 \cdot 10^6 \cdot 6.24 \cdot
10^{18} \cdot 1.78 \cdot 10^{-36} = 3.7 \cdot 10^{-12} \;\; . $$

If the temperature of a flat iron is increased by 200$^0$ its mass
increases by $\Delta m/m = 10^{-12}$. This is readily estimated
using the specific heat (25 J $\cdot$  mol$^{-1}$ K$^{-1} = 450$ J
kg$^{-1}$ K$^{-1}$): $$ \frac{\Delta m}{m} = 450({\rm J ~ kg}^{-1}
{\rm K}^{-1}) 200 \; {\rm K} = 10^{-12} \;\; . $$ All these mass
differences are too tiny to be measured directly. Let us note that
the defect  of mass in a hydrogen atom 13.5 eV is also too small
to be observed directly because the mass of the proton is known
with large uncertainty $\pm 80$ eV.

The tiny values of $\Delta m$ in atomic transitions and chemical
reactions were the basis for the statement that in
non-relativistic physics mass is additive, and of the law of
conservation of this additive mass.

However in nuclear and particle physics the defect of mass is much
larger. For instance, in the case of deutron, which is a nuclear
bound state of proton and neutron, the binding energy and hence
the defect of mass is 2.2 MeV, so that $\Delta m /m \simeq
10^{-3}$.

Of special pedagogical interest is the reaction of annihilation of
electron and positron into two photons (two $\gamma$-quanta).
Photons are massless particles, which always move with velocity
$c$. The latter statement follows from eqs. (\ref{30}),
(\ref{31}), (\ref{34}):
\begin{equation}
|\frac{\rm \bf v}{c}| = |\frac{{\rm\bf p}c}{E}| = 1 \; , \;\; {\rm
if} \;\; |{\rm\bf p} c| = E \;\; . \label{40}
\end{equation}
Depending on their energy, photons are referred to as quanta of
radio waves,  visible and invisible light, $X$-rays,
$\gamma$-quanta.

The reaction of annihilation is
\begin{equation}
e^+ + e^- \to \gamma + \gamma \;\; . \label{41}
\end{equation}

Let us consider the case when electron and positron annihilate at
rest. Then their total  energy is $E=E_0 = 2m_e c^2$, while the
total momentum ${\rm\bf p} =0$. Due to conservation of energy and
momentum the two photons will fly with opposite momenta, so that
each of them will have energy equal to $m_e c^2$. The rest frame
of $e^+ + e^-$ will be obviously the rest frame of two photons.
Thus, the rest energy of the system of two photons will be $2m_e
c^2$ and hence the mass of this system will be $2m_e$, in spite of
the fact that each of the photons is massless. We see that mass in
relativity theory is conserved, but not additive.

In general case the system of two free particles with energies and
momenta $E_1$, ${\rm\bf p}_1$ and $E_2$, ${\rm\bf p}_2$ has total
energy
\begin{equation}
E = E_1 + E_2 \;\; , \label{42}
\end{equation}
and  total momentum
\begin{equation}
{\rm\bf p} = {\rm \bf p}_1 +{\rm\bf p}_2 \;\; . \label{43}
\end{equation}
These equations follow from additivity of energy and momentum. The
mass of the system is defined as before by eq. (\ref{34}). Hence
\begin{equation}
m^2 = (E_1 +E_2)^2 -({\rm\bf p}_1 +{\rm\bf p}_2)^2 = m_1^2 +m_2^2
+ 2E_1 E_2 (1- {\rm\bf v}_1 {\rm\bf v}_2) \;\; . \label{44}
\end{equation}
It follows from eq. (\ref{44}) that the mass of a system of two
particles depends not only on masses and energies of these
particles, but also on the angle between their velocities. Thus
for two photons $m$ is maximal when this angle is $\pi$ and
vanishes when it is zero.

The mass of system has lost its Newtonian meaning of an amount of
substance, its main characteristic being now rest energy (in
units, where $c=1$). Newtonian equations can be obtained from
relativistic ones in the limiting case of low velocities ($v/c\ll
1$). In that case $\gamma$ given by eq. (\ref{32}) becomes
\begin{equation}
\gamma = (1-v^2/c^2)^{-1/2} \simeq 1 + \frac{v^2}{2c^2} \;\; ,
\label{45}
\end{equation}
so that for one particle we get:
\begin{equation}
E = mc^2 +\frac{mv^2}{2} = E_0 + E_{kin} \;\; , \label{46}
\end{equation}
\begin{equation}
{\rm \bf p} \simeq m{\rm\bf v} \;\; . \label{47}
\end{equation}

For a system of two particles in the limit of vanishing $v$ we get
from eq. (\ref{44})
\begin{equation}
m^2 \simeq (m_1 + m_2)^2 . \label{48}
\end{equation}
Thus the approximate additivity of mass is restored.

We started the description of Newtonian mechanics by consideration
of static gravitational and electric interactions, in particular,
their potentials (\ref{8'}) and (\ref{12'}). For particles at rest
these potentials do not depend on time. The situation is
drastically changed when the velocity of particles is not
negligible. Let us start with electrodynamics. First, in addition
to the scalar potential $\varphi$ we have now vector potential
${\rm\bf A}$, so that $\varphi, {\rm\bf A}$ form a four vector.
Second, because of finite velocity $c$ of propagation of
electromagnetic perturbations, both $\varphi$ and ${\rm\bf A}$ are
retarded:
\begin{equation}
\varphi(t_2, {\rm\bf r}_2) = \frac{e}{r-\frac{\rm\bf v r}{c}} \;\;
, \label{49}
\end{equation}
\begin{equation}
{\rm\bf A}(t_2, {\rm\bf r}_2) = \frac{e{\rm\bf
v}}{c(r-\frac{\rm\bf v r}{c})} \;\; , \label{50}
\end{equation}
where
\begin{equation}
{\rm\bf r} = {\rm\bf r}_2(t_1) - {\rm\bf r}_1(t_1) \;\; ,
\label{51}
\end{equation}
while
\begin{equation}
{\rm\bf v} = {\rm\bf v}(t_1, {\rm\bf r}_1) \;\; . \label{52}
\end{equation}

Thus defined $\varphi$ and ${\rm\bf A}$ allow one to calculate the
strengths of electric and magnetic fields:
\begin{equation}
{\rm\bf E} = -\frac{1}{c} \frac{\partial{\rm\bf A}}{\partial t} -
\;{\rm grad} \; \varphi \;\; ,\label{53}
\end{equation}
\begin{equation}
{\rm\bf H} = {\rm rot} \; {\rm\bf A}\;\; , \label{54}
\end{equation}
where the differentiation is performed with respect to $t_2$,
${\rm\bf r}_2$.

In a four-dimensional form the six components of antisymmetric
tensor of the strength of electromagnetic field $F_{ik}$ are
expressed in terms of derivatives of four-dimensional potential:
\begin{equation}
F_{ik} = \frac{\partial A_k}{\partial x^i} - \frac{\partial
A_i}{\partial x^k} \;\; . \label{55}
\end{equation}

The lower indices are referred  to as covariant, while the upper
ones as contravariant.
\begin{equation}
A_i = (\varphi, -{\rm\bf A}) \; , \;\; x^i = (ct, {\rm\bf r}) \; ,
\;\; i=0,1,2,3 \label{56}
\end{equation}

The components of {\bf E} and {\bf H} are expressed in terms of
components of $F_{ik}$:
\begin{equation}
E_x = F_{11} \; , \;\; E_y = F_{02} \; , \;\; E_z = F_{03} \; ,
\label{63}
\end{equation}
\begin{equation}
H_x = F_{23} \; , \;\; H_y = F_{31} \; , \;\; H_z = F_{12} \; .
\label{64}
\end{equation}

The Lorentz invariant products of four-vectors are constructed by
using the so-called metric tensor $\eta_{ik}$, which in an
inertial reference frame is given by a diagonal 4$\times$ 4
matrix:
\begin{equation}
\eta^{ik} = \eta_{ik} = {\rm diag} ({\rm 1, -1, -1, -1})
\label{65}
\end{equation}
with vanishing non-diagonal elements. Multiplication of a
covariant vector by $\eta^{ik}$ gives a contravariant vector,
e.g.:
\begin{equation}
p^i = \eta^{ik} p_k \;\; , \label{66}
\end{equation}
where summation over index $k$ is presumed.

Up to now we considered point-like particles. If the charge is
smeared over a finite volume with density $\rho$, the total charge
of a particle is given by integral:
\begin{equation}
e = \int \rho({\rm\bf r}) d{\rm\bf r} \;\; . \label{67}
\end{equation}
Similarly:
\begin{equation}
e{\rm\bf v} = \int {\rm\bf v}({\rm\bf r}) \rho({\rm\bf r})
d{\rm\bf r} \;\; . \label{68}
\end{equation}
The four-vector
\begin{equation}
j^i = (c\rho, {\rm\bf v}\rho) \label{69}
\end{equation}
describes the density of the four-current. (In the case of a
point-like particle $\rho = e\delta(r-{\rm\bf r}_1)$.)

The famous Maxwell's equations of classical electrodynamics have
the form:
\begin{equation}
\frac{\partial F^{ik}}{\partial x^k} = -\frac{4\pi}{c} j^i \;\; ,
\label{70}
\end{equation}
\begin{equation}
\frac{\partial \tilde{F}^{ik}}{\partial x^k} = 0 \;\; , \label{71}
\end{equation}
Here
\begin{equation}
\tilde{F}^{ik} = \varepsilon^{iklm}F_{lm} \;\; , \label{72}
\end{equation}
where $\varepsilon^{iklm}$ is Lorentz-invariant antisymmetric
tensor. We see that current $j^i$ is the source of electromagnetic
field.

\section{Mass in General Relativity}

Let us now consider relativistic gravity. The role of
gravitational potentials is played by 10 components of symmetric
metric tensor $g_{ik}(x^i)$, four of them being diagonal, while
six off-diagonal. What is very important is that in the case of
gravity the ten components of $g^{ik}$ are functions of space-time
points $x^i$: they change from one point to another. The source of
gravitational field, the analogue of vector $j^i$, is the density
of energy-momentum tensor $T^{ik}$. $T^{ik}$ is symmetric and
conserved
\begin{equation}
\frac{\partial T^{ik}}{\partial x^k} = 0 \;\; . \label{73}
\end{equation}
The total 4-momentum of a system
\begin{equation}
p^i = \frac{1}{c} \int T^{i0}({\rm\bf r}) d {\rm\bf r} \label{74}
\end{equation}
Hence $T^{00}$ is the density of energy, while $T^{10}/c$,
$T^{20}/c$ and $T^{30}/c$ represent the density of momentum. For a
point-like particle with mass $m$ the density of mass $\mu$ is
given by
\begin{equation}
\mu = m\delta ({\rm\bf r} - {\rm\bf r}_1) \;\; . \label{75}
\end{equation}
\begin{equation}
T^{ik} = \mu c \frac{dx^i}{ds} \cdot \frac{dx^k}{dt} = \mu c u^i
u^k \frac{ds}{dt} \;\; , \label{76}
\end{equation}
where $u^i$ is contravariant velocity, while $ds$ is an invariant
interval:
\begin{equation}
u^i = dx^i/ds \label{77}
\end{equation}
\begin{equation}
ds^2 = g_{ik} dx^i dx^k \label{78}
\end{equation}
\begin{equation}
ds = cd\tau = \sqrt{g_{00}} dx^0 \;\; . \label{79}
\end{equation}
Hence
\begin{equation}
\tau = \frac{1}{c}\int\sqrt{g_{00}} dx^0 \;\; , \label{80}
\end{equation}
where $\tau$ is the proper time for a given point in space.

The connection between $u^i$ and ordinary 3-velocity ${\rm\bf v}$
is
\begin{equation}
u^i = (\gamma, \frac{{\rm\bf v}}{c} \gamma) \;\; . \label{81}
\end{equation}
Thus
\begin{equation}
u_i u^i = 1 \;\; . \label{82}
\end{equation}
The most important conclusion is that the source of gravitational
field is proportional to the mass of a particle.

The  equation for gravitational potential $g_{ik}$, derived by
Einstein in 1915, is more complex than the Maxwell equation for
$A^i$:
\begin{equation}
R_{ik} - \frac{1}{2} g_{ik} R = 8\pi GT_{ik} \;\; . \label{83}
\end{equation}
Here $R_{ik}$ is the so-called Ricci tensor, while $R$ is scalar
curvature:
\begin{equation}
R = g^{ik} R_{ik} \;\; . \label{84}
\end{equation}
The role of electromagnetic field strength $F_{ik}$ is  played in
gravity by the affine connection:
\begin{equation}
\Gamma_{kl}^i = \frac{1}{2} g^{im} \left(\frac{\partial
g_{mk}}{\partial x^l} + \frac{\partial g_{ml}}{\partial x^k} -
\frac{\partial g_{kl}}{\partial x^m} \right) \;\; , \label{85}
\end{equation}
while the role of derivative $\partial_k F^{ik}$ is played by the
left-hand side of eq. (\ref{83}), where the Ricci tensor is given
by:
\begin{equation}
R_{ik} = g_{in} g^{rs} \left(\frac{\partial
\Gamma_{ks}^n}{\partial x^r}
-\frac{\partial\Gamma_{kr}^n}{\partial x^s} +\Gamma_{pr}^n
\Gamma_{ks}^p -\Gamma_{ps}^n \Gamma_{kr}^p \right) \;\; .
\label{86}
\end{equation}

The drastic difference of gravidynamics from electrodynamics is
the nonlinearity of Einstein equation (\ref{83}): it contains
products of affine connections. This nonlinearity manifests itself
at low values of $v/c$ as a tiny effect in the precession of
perihelia  of planets (Mercury). However it is very important for
strong gravitational fields in such phenomena as black holes.

From principal point of view of highest priority is the dual role
of the tensor $g_{ik}$, which is both dynamical and geometrical.
Dynamically $g_{ik}$ represents the potential of gravitational
field. On the other hand $g_{ik}$ and its derivatives determine
the geometry of space-time. Einstein gave to his theory of
relativistic gravity the name of General Theory of Relativity
(GTR).

It is clear from the above equations of GTR that in
non-relativistic limit $v/c \ll 1$ the gravitational interaction
is determined by only one mass $m$, while notations $m_g$ and
$m_i$ are redundant and misleading. Even more useless are concepts
of active and passive gravitational mass often considered by some
authors.

\section{The pedagogical virus of ``relativistic mass''}

The ``famous formula $E = mc^2$'' and the concept of
``relativistic mass'' increasing with velocity, which follows from
it, are historical artifacts, contradicting the basic symmetry of
Einstein's Special Relativity, the symmetry of 4-dimensional
space-time. The relation discovered by Einstein is not $E = mc^2$,
but $E_0 = mc^2$, where $E_0$ is the energy of a free body at rest
introduced by Einstein in 1905. The source of the longevity of the
``famous formula'' is the irresponsible attitude of relativity
theory experts to the task of explaining it to the non-experts.

The notion of ``relativistic mass'' presents a kind of pedagogical
virus which very effectively infects new generations of students
and professors and shows no signs of decline. Moreover in the Year
of Physics it threatens to produce a real pandemia.

I published my first articles \cite{11,12} against the ``Einstein
famous equation $E = mc^2$'' in 1989. The subject seemed important
to me because it concerned the proper teaching of special
relativity at high schools, colleges and universities and
explaining its genuine meaning to a wide audience of
non-physicists, the so-called ``pedestrians'' in popular science
magazines and books.

The task looked also not absolutely formidable because a
consistent presentation of relativity existed for a long time in
the world-wide accepted textbook by Landau and Lifshitz \cite{13},
which was the basis of my own understanding, and in some other
textbooks.

The existence of countless texts, in which the essence of
relativity was mutilated (or semi-mutilated) had two sides. On one
hand, it looked discouraging, especially because among the authors
of these texts there were many famous physicists, the fathers and
greatest authorities of modern physics. On the other hand, it was
a challenge. So I tried to  explain clearly to the readers the
beauty of four-dimensional space-time approach and the ugliness
and inconsistency of ``relativistic mass'', an illegitimate child
of relativistic and non-relativistic equations.

My optimism had increased when in 1992 Taylor and Wheeler in the
second edition of the influential and popular ``Spacetime
Physics'' \cite{14} included a ``Dialog: Use and Abuse of the
Concept of Mass'', in which they supported my articles
\cite{11,12}. A copy of this book is in my bookcase with a
postcard sent to me in October 1991 by John Archibald Wheeler. The
postcard has a photo of the famous Albert Einstein Memorial in
front of the building of the National Academy of Sciences,
Washington, DC. The bronze sculpture of Einstein includes a
copybook with $E = mc^2$ on an open page.

Since that time I received hundreds of letters from physicists
(both professors and students) stating their adherence to the
four-dimensional formulation of relativity and to the Lorentz
invariant concept of mass. In a few cases I helped the authors to
correct erroneous explanations of the concept of mass in preparing
new editions of their textbooks. However the number of proponents
of relativistic mass seemed not to decrease.

A leading role in promoting the relativistic mass have played the
books by Max Jammer \cite{15,16}. Especially aggressive  the
proponents of relativistic mass became in connection with the
World Year of Physics, which marks the 100th anniversary of
fundamental articles published by Einstein in 1905.

The campaign started by the September 2004 issue of ``Scientific
American'', full with ``relativistic mass'' equal to $m_0/\sqrt{1
- v^2/c^2}$, where $m_0$ is rest mass, and ``the most famous
equation $E = mc^2$''. A letter to the editors, defending the
four-dimentional approach and invariant mass had been rejected by
the editor G. Collins who in April 2005 wrote: ``Most important,
we believe that tackling the issue head-on in the manner you and
your coauthors want in the letters column of Sci. Am. would be
very confusing to our general audience and it would make the
subject seen all the more mysterious and impenetrable to them''.
Thus to avoid ``head-on'' collision of correct and false arguments
the editors of Sci. Am. preferred to hide from the readers the
correct viewpoint.

P. Rodgers -- the Editor of European ``Physics World'' wrote in
January 2005 in editorial \cite{17}: ``... $E = mc^2$ led to the
remarkable conclusion that mass and energy are one and the same''.
Unlike G. Collins, P. Rodgers published a letter criticizing this
statement and partly agreed with the criticism \cite{18}.

In September 2005 the bandwagon of relativistic mass was joined by
``The New York Times'', which published an article by B. Green
\cite{19}.

The journalists were supported by renowned scientists, such as R.
Penrose, who in a new thousand pages thick book had written
\cite{20}:

``In a clear sense mass and energy become completely equivalent to
one another according to Einstein's most famous equation $E =
mc^2$.''

How many students, teachers and journalists will be infected by
this sentence? How many readers had been infected by the famous
book by S. Hawking \cite{21}, the second edition of which appeared
in 2005? On the very first page of it Hawking wrote:

``Someone told me that each equation I included in the book would
halve the sales. I therefore resolved not to have any equations at
all. In the end, however, I did put in one equation, Einstein's
famous equation $E = mc^2$. I hope that this will not scare off
half of my potential readers.''

I am sure that the usage of $E = mc^2$ had doubled the sales of
his book, the buyers being attracted by the famous brand. But is
it possible to estimate the damage done to their understanding of
relativity theory and to the general level of the literature on
relativity incurred  by this case of spreading the virus?

Two recent preprints by Gary Oas \cite{22, 23} written in the
framework of Educational Program for Gifted Youth at Stanford
University were devoted to the use of relativistic mass. The
author ``urged, once again, that the use of the concept at all
levels to be abandoned'' \cite{22}. The manuscript has been
submitted for publication to the ``Americal Journal of Physics'',
but was rejected as being ``too lengthy'' (it contains 12 pages!).
A lengthy bibliography (on 30 pages) of books referring to special
and/or general relativity is provided in ref. \cite{23} to give a
background for discussions of the historical use of the concept of
relativistic mass. It is easy to forecast the aggressive reaction
of the virus infected community to this attempt to cure it.

\section*{Acknowledgments}

This work is supported by the grant of Russian ministry of
education and science No. 2328.2003.2.


\end{document}